\documentclass[a4paper,12pt]{article}
\usepackage{hyperref}
\usepackage{graphicx}

\newtheorem{theorem}{Theorem}

\title{Emission of autoresonant trajectories and thresholds of resonant pumping.}
\author{O.~M.~Kiselev}

\begin{document}

\maketitle
\begin{abstract}
We study an autoresonant asymptotic behaviour for nonlinear oscillators under slowly changing frequency and amplitude of external driver. As a result we obtain formulas for threshold values of amplitude and frequency of the driver when autoresonant behaviour for the nonlinear oscillator is observed. Also we study a capture into resonance and emission out of the resonance for trajectories of the oscillator. A measure of autoresonant asymptotic behaviours for nonlinear oscillator is obtained.
\end{abstract}

\hspace{3mm}

\noindent{\Large\bf Introduction}

\hspace{3mm}

A phase locking is a well-known phenomenon for wide class of nonlinear oscillators under external driver \cite{PikovskyRosenblumKurth2003}. This phenomenon was suggested for pumping of energy into beams of particles \cite{Veksler1945Eng}, \cite{McMillan1945}. Later was shown that the phase locking and autoresonance can be observed for nonlinear oscillators in general begining from planetary motion \cite{FajansFriedland2001} up to low temperatre technics  \cite{UleyskySosedkoMakarov2010Eng}.

From mathematical point of view, typical equation with autoresonant solutions is a nonlinear oscillator under driver with small amplitude:
\begin{equation}
u''+G(u)=\varepsilon F\cos(\Phi(t,\varepsilon)).
\label{EqForNonlinOscillator}
\end{equation}
Here $G$ is a nonlinear function such that  $G(u)\sim u+g_3 u^3$, as $u\to0$, $\varepsilon$ is a small positive parameter, $f$ is a nonoscillated function of $t$ and $\epsilon$. Without of loss of generality one can see two type of examples. First one is an equation for pendulum with external driver:
\begin{equation}
\frac{d^2 u}{d t^2} +\sin(u)=\epsilon F(t,\epsilon)\cos(t-\Omega(t,\epsilon)),
\label{PerturbedPendulumEq}
\end{equation}
another one is Duffing's oscillator:
\begin{equation}
\frac{d^2 u}{d t^2} +u-\frac{u^3}{6}=\epsilon F(t,\epsilon)\cos(t-\Omega(t,\epsilon)).
\label{DuffingEq}
\end{equation}
Here $0<\epsilon\ll1$, $F(t,\epsilon)$ is an amplitude of perturbation, $\Omega(t,\epsilon)$ is a time shift of the resonant frequency of perturbation. Additional condition on the shift is a slow dependency on $t$. It is convenient to apply: $\Omega'=\epsilon^{2/3}\omega(t,\epsilon)$, where $\omega(t,\epsilon)$ is such that $\omega'(t,\epsilon)=o(1)$. 

A primary resonance equation defines an amplitude of small oscillation of order $O(\sqrt[3]{\varepsilon})$. This equation  has a form:
\begin{equation}
 i\frac{d\Psi}{d\tau}+\left(\omega(\tau) -|\Psi|^2\right)\Psi={\cal F}(\tau)
\label{generatedPrimryResonanceEq}
\end{equation}
Usually this equation is considered for special values of coefficients: 
$\omega\equiv\tau$ and ${\cal F}(\tau)=\hbox{const}$. In such case the solution  has two type of two parametric solutions. First one is a bounded solution with an quadratic grows of frequency. Second one is a growing solution as  $\sqrt{\tau}$. This one is called an autoresonant solution.
\par
One can connect the solutions of (\ref{generatedPrimryResonanceEq}) and solutions of nonlinear oscillator (\ref{EqForNonlinOscillator}). The autoresonant solutions  of (\ref{generatedPrimryResonanceEq}) define nonlinear oscillations of (\ref{EqForNonlinOscillator}) with growing amplitude. Numerous applications of (\ref{EqForNonlinOscillator}) one can see, for example in  \cite{FajansFriedland2001}. Threshold values of parameters for autoresonance were discussed in  \cite{FriedlanScholarpedia}. Asymptotical analisys for the threshold value for external force for two coupled oscillators was done in \cite{GlebovKiselevLazarev2007Eng}. Asymptotic behaviour for solutions over long time were considered in review \cite{Kalyakin2008}.
\par
For the autoresonant grow is important asymptotic behaviour of $\omega(\tau)$ and ${\cal F}(\tau)$ in  (\ref{generatedPrimryResonanceEq}). Below we consider the coefficients of algebraic type: $\omega(\tau)=\lambda^2\tau^{2a}$ and ${\cal F}(\tau)=f\tau^b$, where $a>0,\,b,\lambda,f\in \mathbf{R}$.

\par
In this work we find intervals for values of  amplitude and frequency of driver which contain autoresonant behaviour of solution. In particular we found that the stable autoresonant growth exists for decreasing amplitudes of driver. We study stability of autoresonant behaviour and show  values  of parameters for stability, see also \cite{GlebovKiselevTarkhanov2017}. This result is more general that earlier studies of autoresonance for special case  ${\cal F}\equiv const$ and $a=1/2$ which are discussed in \cite{Kiselev2006Eng}, \cite{KalyakinSultanov2013Eng}, \cite{Sultanov2014Eng}. 

\par 
Here we obtain results about capture and emission of trajevtories concerning the autoresonance area. This approach is extension of earlier short preprint \cite{Kiselev2013ArXiv}. We should remind earlier resuts on this field \cite{Neish1975Eng}, \cite{Haberman1977} and more complete asymptotic approach in special case \cite{KiselevGlebov2003}.

\par
Let us discuss general structure of this work. In the section \ref{secGenesys} we give known approach to derive equation for primary autoresonance for nonlinear oscillaors under external driver. In the section \ref{secThresholdValues} we have found necessary conditions for parameters of equation (\ref{generatedPrimryResonanceEq}) when  growing autoresonant behaviour may exist. In the section  \ref{secAsymptoticSubstitution} we derive the equation for autoresonance and obtain slow varying autoresonant asymptotics.  In the section  \ref{secCapture} we show an equation which define a capture into autoresonance and  calculate a gap between separatrises and  get a measure for trajectories which pass though this gap. In the section \ref{secAsymptoticsOfMesure} we obtain an asymptotic behaviour of a measure for autoresonant solutions of nonlinear oscillator.

\section{Genesis of the primary autoresonant equation}
\label{secGenesys}

The equation for primary autoresonant defines a long temporary evolution of amplitude for fast oscillations of nonlinear oscillator under resonant perturbation with slow shift of the frequency of perturbation. As an example one can consider equation for perturbed pendulum  (\ref{PerturbedPendulumEq}) or equation for Duffing's oscillator (\ref{DuffingEq}). One can find the solutions of these equation for small values of $u$ in form of fast oscillating function with slow modulated amplitude:
$$
u=\sqrt[3]{\epsilon} A e^{i(t-\Omega)}+c.c..
$$
Here $A=A(\tau)$ and  $\tau=\epsilon^{2/3} t$ is slow time. Let us substitute this formula into  (\ref{PerturbedPendulumEq}) then one get:
$$
\epsilon\left(2i A'+2\omega A-\frac{1}{2}|A|^2 A\right)e^{i(t-\Omega)}+o(\epsilon)= \frac{1}{2}Fe^{i(t-\Omega)}.
$$
Then in primary order on $\epsilon$ one gets the equation for slow varying amplitude 
$A(\tau)$. 
$$
2i A'+2\omega A-\frac{1}{2}|A|^2 A=\frac{1}{2}F.
$$
Let us change $A=2\psi$ and $F=8{\cal F}$ then we get equation  (\ref{generatedPrimryResonanceEq}) for complex-valued function $\psi$.

Such approach for a derivation of autoresonant equation is followed by averaging method  \cite{KrylovBogolyubov1937Eng}. This approach for autoresonance is well-known,  see examples \cite{Friedland2000PhysRev},\cite{Kalyakin2002Eng}. The detalled review for autoresonace and its connection with nonlinear oscillators one can get in monograph \cite{GlebovKiselevTarkhanov2017}. Asymptotic analysis for autoresonant growth for nonlinear oscillator from small up to $O(1)$ is developed in  \cite{Garifullin2003Eng}. 

\section{Threshold values of order for growth of frequency and amplitude}
\label{secThresholdValues}
Let us substitute coefficients into (\ref{generatedPrimryResonanceEq}):
\begin{equation}
i\frac{d\Psi}{d\tau}+\left(\lambda^2\tau^{2a} -|\Psi|^2\right)\Psi=f\tau^b
\label{primryResonanceEq}
\end{equation}
The amplitude of leading-order term of autoresonant asymptotic behaviour for solution as $\tau\to\infty$ is defined from the following equation:
$$
\left(\lambda^2\tau^{2a} -|\Psi|^2\right)=o(\tau^{2a}).
$$
Therefore for autoresonant solutions one gets  a condition: $|\psi|\sim \lambda\tau^a$. 

It is convenient to write equation (\ref{primryResonanceEq}) in such form that the growing part of solution can be obvious allocated. Let us change:
$$
\Psi=\lambda\tau^a\, \psi\, \exp\left(i\lambda^2\frac{\tau^{2a+1}}{2a+1}\right).
$$
As a result one get:
$$
i\tau^{-2a}\lambda^{-2}\psi' -|\psi|^2\psi 
=\frac{f\tau^b}{\lambda^3\tau^{3a}}\exp\left(-i\lambda^2\frac{\tau^{2a+1}}{2a+1}\right)-i\frac{a\psi}{\lambda^2 \tau^{2a+1}}.
$$
Here it is convenient to use new variable : $\theta=\lambda^2\tau^{2a+1}/(2a+1)$ and assume that  $\psi=\psi(\theta)$. Then the equation for primary resonance has a form::
$$
 i\frac{d\psi}{d\theta}-|\psi|^2\psi=
 \frac{(2a+1)^{\frac{b-3a}{2a+1}}}{\lambda^{\frac{3+2b}{2a+1}}}
 \frac{f\,\exp\left(-i\theta\right)}{\theta^{{\frac{3a-b}{2a+1}}}}\,-
 \frac{ia}{2a+1}\ \frac{\psi}{\theta}.
$$
Let us define:
$$
\nu=\frac{3a-b}{2a+1},\quad h=f (2a+1)^{-\nu} \lambda^{-\frac{2b+3}{2a+1}},\quad q=\frac{a}{2a+1}.
$$
As a result we obtain the equation for $\psi$ in the form:
\begin{equation}
i\frac{d \psi}{d \theta}-|\psi|^2 \psi =\frac{h e^{-i\theta}}{\theta^\nu}-i\frac{q \psi}{\theta}.
\label{primaryResonanceEqInPerturbedForm}
\end{equation}
If $\nu>0$, then the right-hand side of the equation is small as  $\theta\to\infty$. To study an asymptotic behaviour of solutions for large $\theta$ we obtain an equation for $|\psi|^2$. For this we use equations for  $\psi$ and complex gonjugation $\overline{\psi}$:
\begin{eqnarray*}
\frac{d \psi}{d \theta}=
&
-i\frac{h e^{-i\theta}}{\theta^\nu}-\frac{q \psi}{\theta},
\\
\frac{d \overline{\psi}}{d \theta}=
&
i\frac{h e^{i\theta}}{\theta^\nu}-\frac{q \overline{\psi}}{\theta}.
\end{eqnarray*}

Let us differentiate  $|\psi|^2$ using these equations:
\begin{equation}
\frac{d |\psi|^2}{d \theta}=\frac{h}{\theta^\nu}(i\psi e^{i\theta}-i\overline{\psi}e^{-i\theta})-
2\frac{q}{\theta}|\psi|^2.
\label{squaredPsiEq}
\end{equation}
Here we define $\psi=r e^{i\phi}$, where $r=r(\theta)$ and $\phi=\phi(\theta)$ are real-valued functions. Then equation (\ref{squaredPsiEq}) can be rewrited in the form of differential equation for $r$:
$$
\frac{d r}{d\theta}=-\frac{h}{\theta^\nu}\sin(\theta+\phi)-\frac{q}{\theta}r.
$$
Let us find maximum of order for growth of  $r$. For this we estimate: $\sin(\theta+\phi)=-1$.  Let us define a majorant $R>r$. For this majorant one gets  an equation:
$$
\frac{d R}{d\theta}=\frac{h}{\theta^\nu}-\frac{q}{\theta}R.
$$
Then
$$
R=\frac{\theta^{1-\nu}}{1+q-\nu}+C\theta^{-q},\quad C\in\mathbf{R}.
$$
Using formula for $q$ one gets   $0<q<1/2$ as $a>0$. The majorant does not decrease if $\nu<1$. As a result one gets $0<\nu<1$. Therefore the parameters  $a,b$ are:
$$
0<\frac{3a-b}{2a+1}<1.
$$
 
\begin{theorem}
The necessary conditions for existing of decreasingless solution are:
$$
a>0,\quad a-1\le b\le 3a.
$$ 
\end{theorem} 

For linear chirp-rate the frequency  $\omega$ defines by  $a=1/2$. Then the necessary conditions for autoresonanse for the amplitude of the driver are $\tilde f_1\tau^{-1/2}\le f\le \hat f_1\tau^{3/2}$, where $\tilde f_1,\hat f_1$ are some constants.

\section{Asymptotic substitution}
\label{secAsymptoticSubstitution}

To find asymptotic behaviour for solutions of (\ref{primaryResonanceEqInPerturbedForm}) as $\theta\to\infty$ one can write equation for amplitude and phase for function $\psi$. Let us change:
\begin{equation}
\psi=(1+\rho) e^{i(\alpha-\theta)},
\label{newFormOfSolution}
\end{equation}
here $\rho,\alpha$ are real valued functions of $\theta$.

Let us substitute these formulas into  (\ref{primaryResonanceEqInPerturbedForm}), multiply the left and right-hand sides of the equation by  $e^{-i(\alpha-\theta)}$. Then we obtain system of equations for real and imaginary parts of (\ref{primaryResonanceEqInPerturbedForm}):
\begin{eqnarray}
\alpha'+\frac{h}{\theta^{\nu}}\frac{\cos(\alpha)}{(1+\rho)}+\rho^2+2\rho=0,
\nonumber\\
\rho'+\frac{h}{\theta^\nu}\sin(\alpha)+\frac{q}{\theta}\rho+\frac{q}{\theta}=0.
\label{alphaRhoEq}
\end{eqnarray}
Here one should find values of coefficients of this system, for which the necessary condition of existing autoresonant solutions are valid: $0<\nu<1$. 

To construct a solution with small $\rho$ it is convenient to rewrite:$\sigma(\theta)=\sin(\alpha)$, тогда $\sigma'=\alpha'\cos(\alpha)$. The system can be written as follows:
\begin{eqnarray}
\sigma'+\frac{h}{\theta^{\nu}}\frac{1-\sigma^2}{(1+\rho)}+(\rho^2+2\rho)\kappa\sqrt{1-\sigma^2}=0,
\nonumber\\
\rho'+\frac{h}{\theta^\nu}\sigma+\frac{q}{\theta}\rho+\frac{q}{\theta}=0.
\end{eqnarray}
Here $\kappa=\hbox{sign}(\cos(\alpha))$ .

\subsection{Slow varying asymptotic expansions}
\label{subsecSlowlyCHanchingSolutions}
When $\theta\to\infty$ the system of equations has two small parameters. There are $\theta^{\nu-1}$ and $\theta^{-\nu}$. An asymptotic expansions for formal series one can obtain using two scaling method, where $\zeta=\theta^{\nu}$, $\xi=\theta^{1-\nu}$,

$$
\frac{d}{d\theta}=\frac{d \zeta}{d\theta}\frac{\partial}{\xi\partial\zeta}+\frac{d \xi}{d\theta}\frac{\partial}{\zeta\partial\xi}
=\frac{\nu}{\xi}\frac{\partial}{\partial\zeta}+\frac{1-\nu}{\zeta}\frac{\partial}{\partial\xi}.
$$ 

The system of equations can be rewritten as follows:
\begin{eqnarray}
\frac{\nu}{\xi}\frac{\partial\sigma}{\partial\zeta}+\frac{1-\nu}{\zeta}\frac{\partial\sigma}{\partial\xi}+
\frac{h}{\theta^{\nu}}\frac{1-\sigma^2}{(1+\rho)}+(\rho^2+2\rho)\kappa\sqrt{1-\sigma^2}=0
\nonumber
\\
\frac{\nu}{\xi}\frac{\partial\rho}{\partial\zeta}+\frac{1-\nu}{\zeta}\frac{\partial\rho}{\partial\xi}+
\frac{h}{\zeta}\sigma+\frac{q}{\zeta\xi}\rho+\frac{q}{\zeta\xi}=0.
\label{Eq-for-two-scale-method}
\end{eqnarray}
Formal solution for this system as a series of negative powers of $\xi$ and $\zeta$ has a form:
$$
\sigma\sim \sum_{j,k>0} \frac{\sigma_{jk}}{\xi^j\zeta^k},\quad \rho\sim\sum_{j,k>0} \frac{\rho_{jk}}{\xi^j\zeta^k},\quad  \sigma_{jk},\rho_{jk}\in\mathbf{R}.
$$
The coefficients of the asymptotic expansions for  $\sigma_{jk}$ and $\rho_{jk}$ are obtained using recurrent equations for orders $\xi^{-j}\zeta^{-k}$.

Leadin-order terms have a form:
$$
\sigma\sim-\kappa\frac{q}{h\xi},\quad \rho\sim -\kappa\frac{h}{2\zeta}.
$$

\par
Using system of equations for  $\rho$ and  $\alpha$  and the formulas for leading-order terms  we obtain: 
\begin{theorem}
Asymptotic behaviour of decreasing solution for $(\alpha_*,\rho_*)$ as $\theta\to\infty$ has a form:
\begin{eqnarray}
 \alpha_{*}\sim \alpha_0+(-1)^{n+1}\frac{q}{h\theta^{1-\nu}},\quad 
\rho_{*}\sim (-1)^{n+1}\frac{h}{2\theta^{\nu}}-\frac{3h^2}{8\theta^{2\nu}},
\label{asymptoticsOfAutoresonance}
\\
\quad \alpha_0=\pi n,\,n=0,1\nonumber.
\end{eqnarray}
\end{theorem}

Here we obtain pair of particular asymptotic solutions. These solutions do not contain parametes and one can consider that these solutions are slow varying equilibriums for nonautonomous system of equations (\ref{Eq-for-two-scale-method}).

\subsection{Fast oscillated asymptotic expansions}
\label{secFastOscillatedAsymptoticExpansions}

\begin{figure}
\includegraphics[scale=0.55]{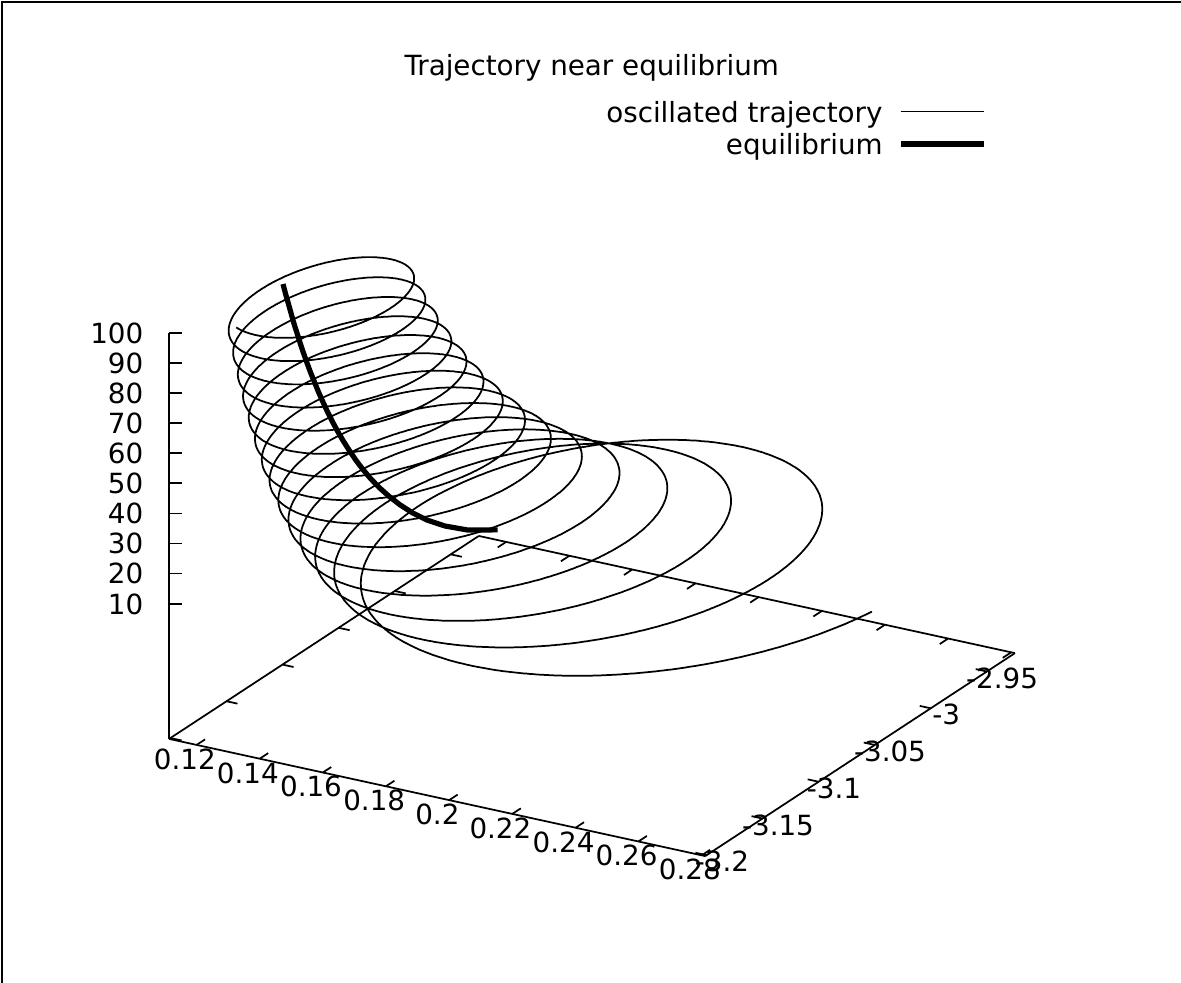}
\includegraphics[scale=0.55]{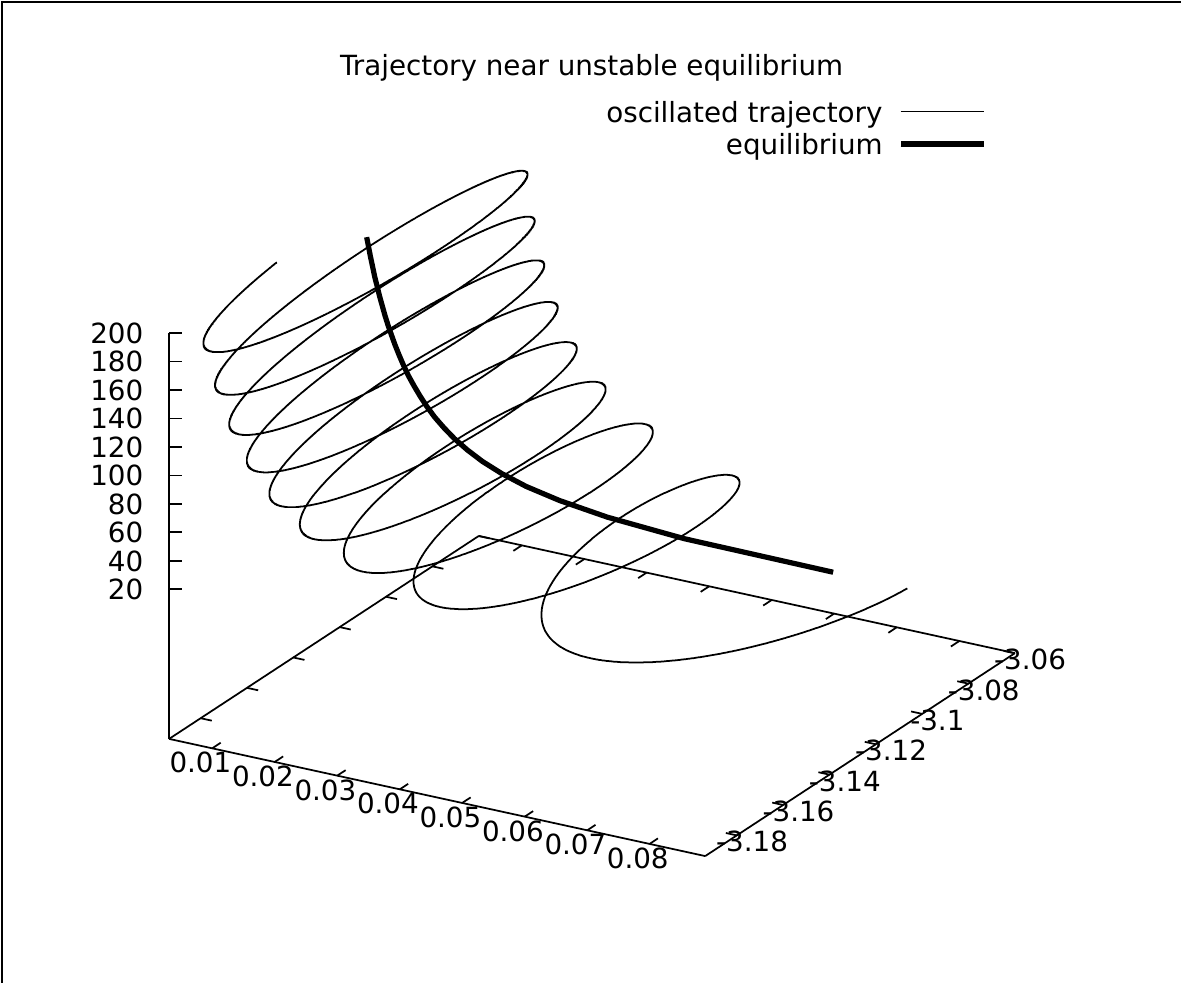}
\caption{Here one can see two oscillated trajectories near slow varying equilibrium. The trajectory was obtained by numerical solution of Cauchy problem with $\alpha=-\pi+q/(h\theta_0^{1-\nu}),\rho=h/(2\theta_0^\nu)$, as $\theta_0=10$.  The numeric solution of Cuchy problem for (\ref{alphaRhoEq}) was integrated by Runge-Kutt forth order method with step  $0.01$. The trajectory near stable slow varying equilibrium is shown on the left picture. Parameters of (\ref{alphaRhoEq}) are $\nu=1/4,q=1/2,h=1$. The trajectory near unstable equilibrium is shown on the right picture. Parameters of (\ref{alphaRhoEq}) are $\nu=3/4,q=1/16,h=1$.}
\label{figOscillatedTrajectory}
\end{figure}
In neighbourhoods of slow varying asymptotic solutions one can construct general solutions which contain arbitrary parameters. One can see such trajectory on figure \ref{figOscillatedTrajectory}. Such solutions depend on fast variable as $\theta\to\infty$. The dependency of the fast variable we separate using system of equations in form  (\ref{alphaRhoEq}).

Let us define: 
$$
\alpha=\alpha(S),\quad \rho=\theta^{-\nu/2}r(S), \quad S=\frac{2}{2-\nu}S^{1-\nu/2}.
$$ 
Here $S$ is new fast variable and $r(S)$ represents function $\rho$ in new scale. It yields:
\begin{eqnarray}
\frac{d\alpha}{d S}=-2r-\theta^{-\nu/2}\left(r^2 +\frac{h\cos(\alpha)}{1+\theta^{-\nu/2}r}\right),
\nonumber\\
\frac{d r}{dS}=-h\sin(\alpha)-\theta^{-1+\nu}q-\theta^{-1+\nu/2}\left(q-\frac{\nu}{2}\right)r.
\label{SystemForPerturbedPendulum}
\end{eqnarray}
One can consider this system as system of equations in closed form, if one rewrite the variable  $\theta$ through new variable  $S$: 
$$
\frac{dS}{d\theta}=\theta^\nu/2,\quad \theta=\frac{2}{2-\nu}\left(\frac{2-\nu}{2}\right)^\frac{2}{2-\nu}S^\frac{2}{2-\nu}.
$$
The formula for $\theta$ of  $S$ looks to large therefore we will use $\theta$ and will keep in mind that it is a function on $S$.

We rewrite  nonautonomous system (\ref{SystemForPerturbedPendulum}) into form of  a perturbed equation for  pendulum. For this one should differentiate on $S$ first equation of the system using second equation of the system. As a result one gets differential equation of second order. Then one should change $r(S)$ by root of algebraic equation of third order for $r$:
$$ 
r=-\frac{1}{2}\frac{d\alpha}{d S}-\frac{1}{2}\theta^{-\nu/2}\left(r^2 +\frac{h\cos(\alpha)}{1+\theta^{-\nu/2}r}\right),
$$
Asymptotic behaviour of smallest root as  $\theta\to\infty$ has a form:
$$
r\sim -  {{{{d}\over{d\,S
 }}\,\alpha}\over{2}} -{{\left({{d}\over{d\,S}}\,\alpha\right)^2+4\,h\,\cos(\alpha)}\over{8\,\theta^{{{\nu
 }\over{2}}}}}+{{-\left(({{d}\over{d\,S}}\,\alpha\right)^3-8\,h\,\cos(\alpha)\,
 \left({{d}\over{d\,S}}\,\alpha\right)}\over{16\,\theta^{\nu}}}.
$$
Full form of equation in form of perturbed pendulum, which is obtained from (\ref{SystemForPerturbedPendulum}), is too long. Therefore we get asymptotic form of this equation and we keep such  terms, which are important for our study as  $\theta\to\infty$.
\begin{eqnarray}
{{d^2 \alpha}\over{d\,S^2}}\sim
&
2\,h\,\sin(\alpha)+2\,q\,\theta^{\nu-1}-\frac{1}{2}{{\,\,\theta^{{{\nu}\over{2}}-1}
 \,\left(4\,q-\nu\right)}}{{d \alpha}\over{d\,S}}
\nonumber
\\
& 
-\theta^{-1}{{2\,h\,q\,\cos(\alpha)}} 
-\theta^{-{{\nu}\over{2}}-1} h\,q\,\cos(\alpha)\,
\left({{ d \alpha}\over{d\,S}}\,\right).
\label{EqInPerturbedPendulumFormAsymptotic}
\end{eqnarray} 
On bounded intervals of $S$ the function $\alpha$ is close to solution  of equation for nonperturbed pendulum. In particular, the slow varying asymptotic expansions are close to equilibriums of unperturbed pendulum were constructed in the section  \ref{subsecSlowlyCHanchingSolutions}. 

In a projection of phase space of equation  (\ref{EqInPerturbedPendulumFormAsymptotic}) on the plane $(\alpha,(d\alpha)/(d S))$ the slow varying solution near  point  $(0,0)$ looks as a saddle due to structural stability of the suddle inder perturbations. 
  
The slow varying solution near point $(\pi,0)$ looks like a center. But it is well-known that the center in unstable under perturbations and it can be transformed into focus. Therefore we should study properties of the solutions near slow varying equilibrium which is close to $(\pi,0)$.

Let us consider an energy of unperturbed pendulum:
$$
E=\frac{(\alpha')^2}{2}+2h\cos(\alpha).
$$
This term for perturbed equation is a function on $S$. Let us derive a derivative of this function under perturbation  (\ref{EqInPerturbedPendulumFormAsymptotic}):
\begin{eqnarray*}
\frac{d E}{d S}
\sim 
2\,q\,\theta^{\nu-1}{{d \alpha}\over{d\,S}}-\frac{1}{2}{{\,\,\theta^{{{\nu}\over{2}}-1}
 \,\left(4\,q-\nu\right)}}\left({{d \alpha}\over{d\,S}}\right)^2
\nonumber
\\ 
-\theta^{-1}{{2\,h\,q\,\cos(\alpha)}} {{d \alpha}\over{d\,S}}
-\theta^{-{{\nu}\over{2}}-1} h\,q\,\cos(\alpha)\,
\left({{ d \alpha}\over{d\,S}}\,\right)^2.
\end{eqnarray*}
The changing of $E$ over one oscillation is:
$$
\oint dE\sim  -\frac{1}{2}{{\,\,\theta^{{{\nu}\over{2}}-1}
 \,\left(4\,q-\nu\right)}}\oint\left({{d \alpha}\over{d\,S}}\right)d\alpha.
$$
So a growth or an increase of $E$ under oscillation of $\alpha$ is defined by a sign of  $4q-\nu$. In the terms of parameters for equation (\ref{primryResonanceEq}) one gets a condition $a+b\not=0$.  Therefore we get:

\begin{theorem}  
If $a+b\not=0$, then $E$ is change under one oscillation on the value, which is proportional by area inside the curve on plane $(\alpha,(d\alpha)/(dS))$. 
\end{theorem}
Corollary. If  $a+b>0$, then the slow varying equilibrium near $(\pi,0)$ on plane $(\alpha,(d\alpha)/(dS))$ is stable.

\section{Capture into autoresonance and emission of trajectories from resonant area}
\label{secCapture}

\begin{figure}
\includegraphics[scale=0.4]{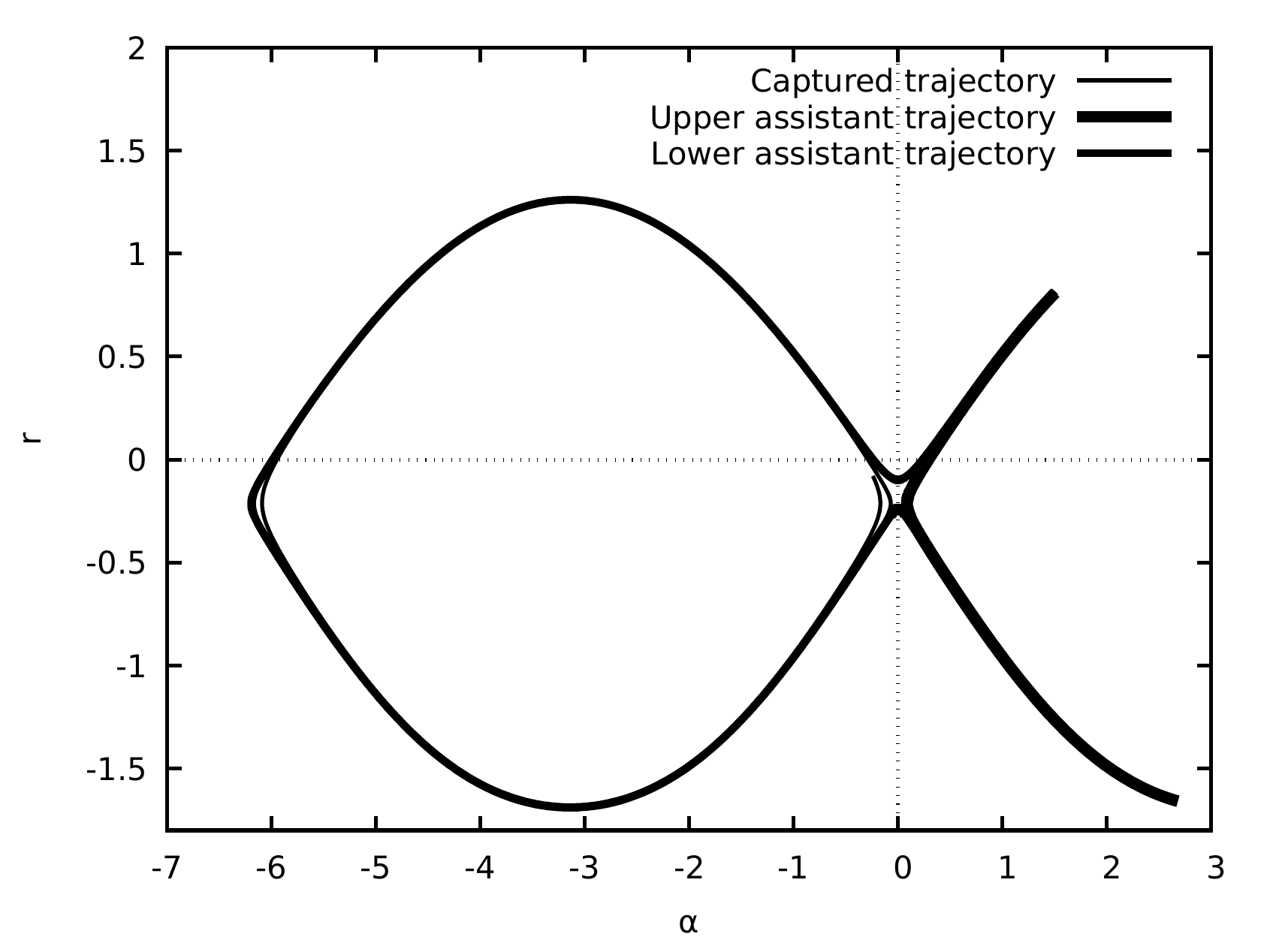}
\includegraphics[scale=0.4]{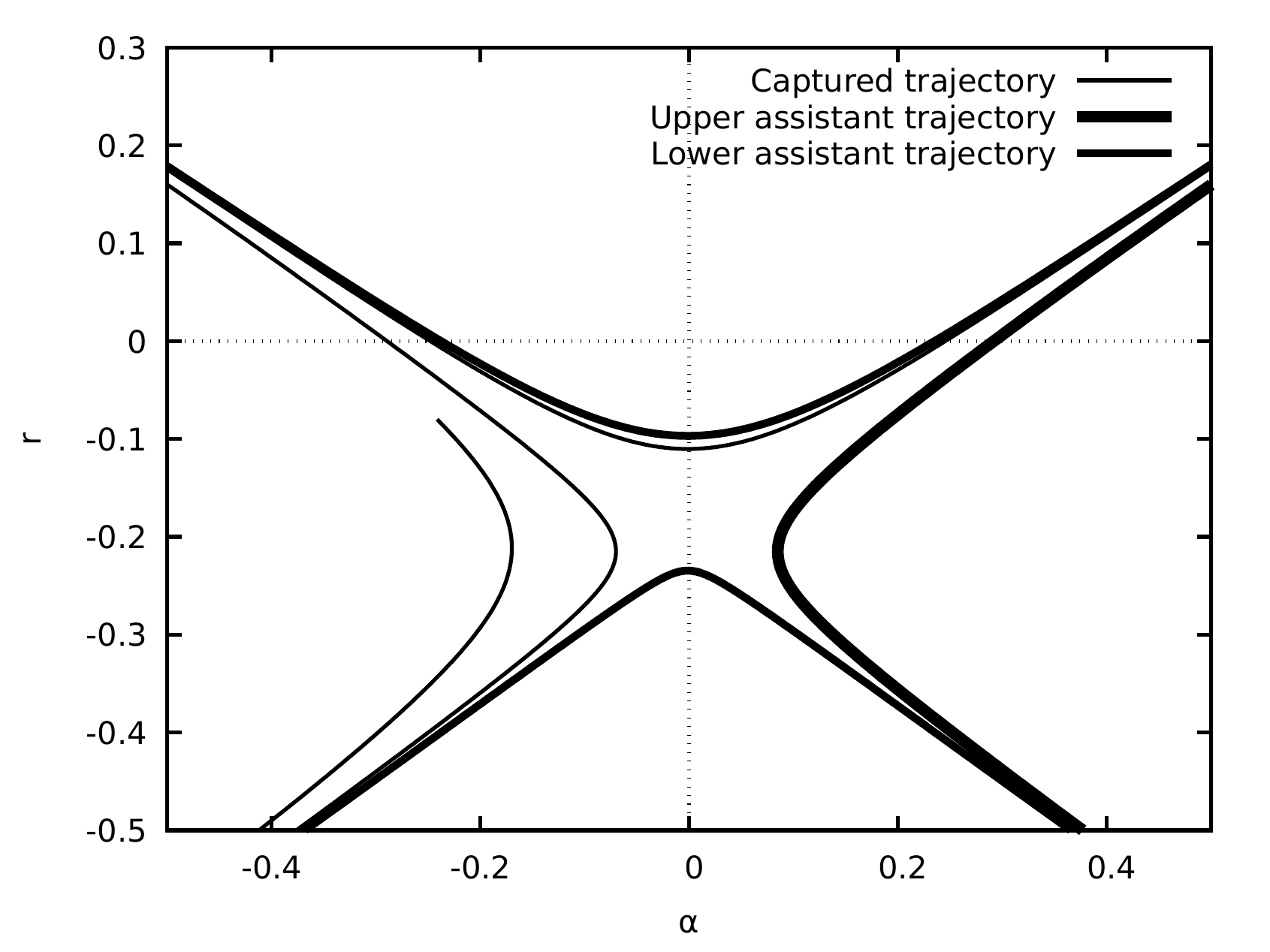}
\caption{Here we show projections on plane $\alpha,r$ for trajectories of system (\ref{SystemForPerturbedPendulum}), which are obtained by Runge-Kutt forth order method with step  $0.01$. Values of parameters for the system are $h=1,\,q=1/2,\,nu=1/4$. For the captured trajectory the solution was obtained by Cauchy problem in inverse direction of  $S$  from $S=992$ to $S=1020$ and Cauchy data were: $S=992,\,\alpha=-\pi+2.9,\,r=-0.08$. For assistant trajectories the solutions were obtained by solution of Cauchy problems for positive direction  by $S$. The solution with largest energy we obtain  on the interval $S\in[992,1010]$ with Cauchy data  $r=0.0815,\,\alpha=1.505,\,S=992$. For solution with smallest energy we obtain on the interval $S\in[992,998]$ with Cauchy data: $r=0.0810,\,\alpha=1.505,\,S=992$.}
\label{FigForCapturedTrajectories}
\end{figure}

One can divide  trajectories of the system  (\ref{SystemForPerturbedPendulum}) on resonant trajectories and assistant trajectory. Such trajectories one can obtain by numerically using Runge-Kutt method for Cauchy problem with different parameters for large independent variable    $S$. In particular fig. \ref{FigForCapturedTrajectories} shows  results of such numeric solutions. There are three trajectories with close behaviour far from saddle. But near the saddle the assistant trajectory with largest energy goes through loop. The assistant trajectory with lowest energy make a turn and third trajectory is captured into the resonance area inside the loop.  

\begin{figure}
\includegraphics[scale=0.4]{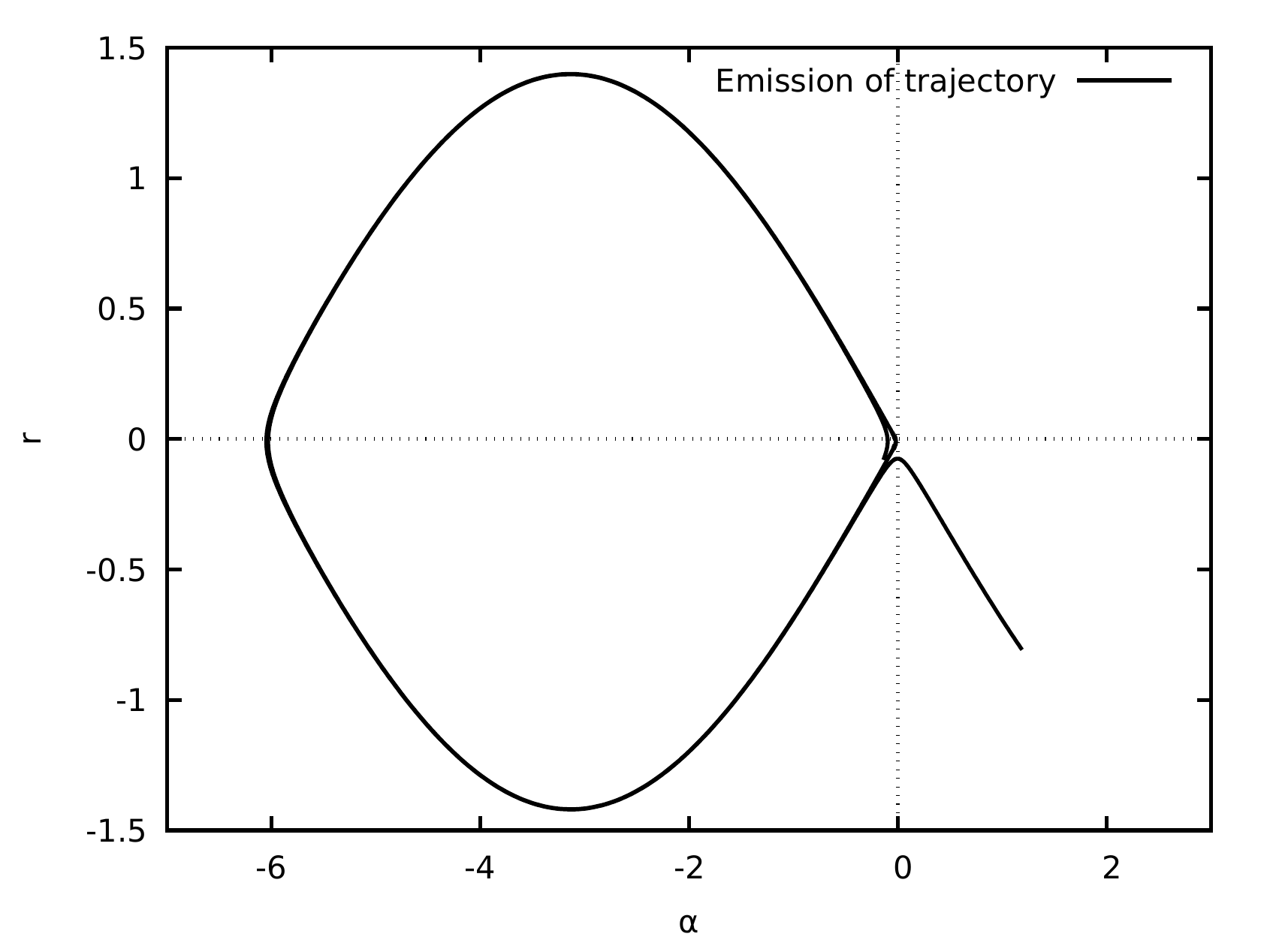}
\includegraphics[scale=0.4]{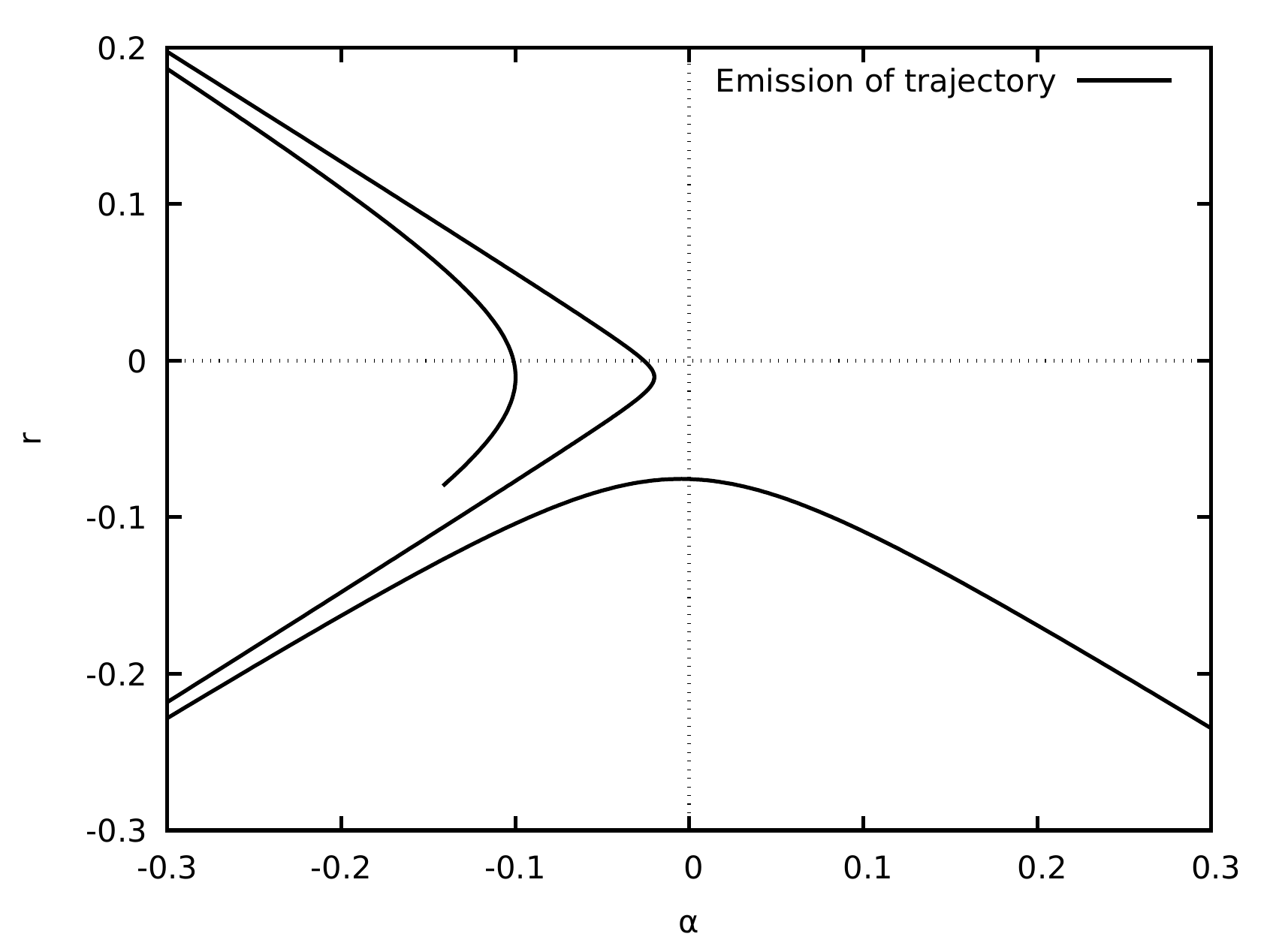}
\caption{Here one can see the projection on plane $\alpha,r$ of trajectories for the system  (\ref{SystemForPerturbedPendulum}). The trajectories were obtained by Runge-Kutt forth order method with step $0.01$. The values of parameters for the system are $h=1,\,q=1/16,\,\nu=3/4$. This trajectory goes away from the resonant area. This shows the emission of the trajectory from resonance. The solution was obtained on the interval $S\in[992,1020]$ with Cauchy data: $S=992,\,\alpha=-\pi+3.0,\,r=-0.08$.}
\label{FigForEmissedTrajectories}
\end{figure}

Depending on the parameters for the system of equations  (\ref{SystemForPerturbedPendulum}) one can obtain not only  trajectories, which are captured into the resonance area, but also one can see an emission of trajectories from the resonant area. Such emission of trajectories was not observed by author elsewhere  for the autoresonant equations. One can see the emissed trajectory on the fig. \ref{FigForEmissedTrajectories}.

Let us consider a behaviour of solutions of (\ref{EqInPerturbedPendulumFormAsymptotic}) near slow varying equilibriums which looks as a saddle on the plane $(\alpha,(d\alpha)/(dS))$.  One can consider equation  (\ref{EqInPerturbedPendulumFormAsymptotic}) as an equation for pendulum with small momentum $M$:
\begin{equation}
\frac{d^2 a}{dS^2}=2h\sin(a)+M.
\label{EqPendulumWithMomentum}
\end{equation}
Here the momentum is: $M\sim 2q\theta^{\nu-1}$ as $\theta\to\infty$. The equation with constant external momentum is integrated by quadrature. In general case the equation with external momentum is not integrable. However in our case the dependency of time is slow with respect to frequency of oscillations. Formally this gives opportunity to consider the equation with slow varying momentum as a perturbation of equation with constant momentum.

There are three type of separatrices of unperturbed pendulum with constant momentum. First one are a homoclinic  loops which are started and finished in the saddles. Second one and third one  are the unbounded separatrices  which are finished or started from the saddle. Due to Melnikov's theory the centers and homoclinics are destroyed under a perturbation   \cite{Melnikov1963Eng}. As a result instead of the separatrix loops one obtain the curve which are go out from saddles to focuses or otherwise.

The trajectories, which are captured into resonance, pass through the gap between two separatrix closeto  the saddle, see \cite{Kiselev-Tarkhanov2014Chaos}, \cite{KiselevTarkhanov2014JMP}.

An important parameter for the pendulum is an energy:
$$
E=\frac{(\alpha')^2}{2}+h\cos(\alpha)-M\alpha.
$$
The  change of the energy near the saddle point define the separatrix gap, which appears due to the perturbation \cite{Melnikov1963Eng}. For equation  (\ref{EqInPerturbedPendulumFormAsymptotic}) the value of the gap between two separatrices is asymptotically close to the change of energy over one oscillation near the separatrix loop of unpertubed penduluum:
$$
\Delta E\sim-\frac{1}{2}{{\,\,\theta^{{{\nu}\over{2}}-1}
 \,\left(4\,q-\nu\right)}}\oint_{\mathcal L}\left({{d \alpha}\over{d\,S}}\right)d\alpha.
$$
Here the curve $\mathcal{L}$ is a separatrix loop for  (\ref{EqPendulumWithMomentum}) when  $M=2q\theta^{\nu-1}$. The integral defines a square inside of this loop. Due to small value of the external momentum this square is asymptotically close to the square of oscillation region of phase space for the pendulum when $M\equiv0$. Then we obtain:
$$
\oint_{\mathcal L}\left({{d \alpha}\over{d\,S}}\right)d\alpha\sim4\sqrt{h}\int_{0}^{2\pi}\sqrt{1-\cos(\alpha)}d\alpha.
$$
The direct calculation shows:
$$
\oint_{\mathcal L}\left({{d \alpha}\over{d\,S}}\right)d\alpha\sim16\sqrt{2h}.
$$
This means that the gap between the separatrices is:
$$
\Delta E\sim8\,\,\theta^{{{\nu}\over{2}}-1}(\nu-4q)\sqrt{2h}.
$$
If  $\nu-4q\not=0$, then  the trajectories are captured into the resonance when $\nu-4q<0$, otherwise the trajectories go out the resonance when  $\nu-4q>0$. Therefore as $\nu-4q>0$ one can see the emission of the trajectories from the resonant area. 

If one rewrite the values of parameters  $q$ and $\nu$ using  $a$ and $b$ from primary equation (\ref{generatedPrimryResonanceEq}) then one gets.

\begin{theorem}
Let the parameters $a$ and $b$ of (\ref{generatedPrimryResonanceEq}) be such that  $a+b\not=0$. Then the trajectories are captured into the resonance for $a-b>0$ and are went out  from the resonance for $a-b<0$.
\end{theorem}

\section{Asymptotic measure of autoresonant solutions}
\label{secAsymptoticsOfMesure}

Let us calculate a measure of trajectories of  (\ref{PerturbedPendulumEq}), which are into autoresonance. This calculations use the constructions of sections  \ref{secGenesys} and \ref{secCapture}. In particular the constructed asymptotics are connected with nonlinear oscillator under external pumping by following formulas:
\begin{eqnarray*}
u(t,\epsilon)\sim\epsilon^{1/3}2\lambda\tau^a(1+\theta^{-\nu/2}r)\exp(i\alpha)\exp\left(i(t-\Omega)\right)+c.c..
\\
\frac{d}{d t} u(t,\epsilon)\sim \epsilon^{1/3}2i(1-\epsilon^{2/3}\lambda^2\tau^{2a})2\lambda\tau^a(1+\theta^{-\nu/2}r)\exp(i\alpha)\exp\left(i(t-\Omega)\right)+c.c..
\end{eqnarray*}
The asymptotic behaviour of the measure in plane cross-section for given $t$ of the phase space $(u,u_t,t)$ is:
$$
Mes=\int_{\mathcal L} d u_t\wedge du \sim \epsilon^{2/3}4\lambda^2\tau^{2a}\theta^{-\nu/2}
\int_{\mathcal L} dr\wedge d\alpha\sim \epsilon^{2/3}4\lambda^2\tau^{2a}\theta^{-\nu/2}
\oint_{\mathcal L} r d\alpha.
$$
The asymptotic behaviour of the square inside the separatrix loop $\mathcal L$ is calculated in the section \ref{secCapture}. One should change the variables $\tau$ and $\theta$ on to the variable $t$ and parameter $\epsilon$:
\begin{eqnarray*}
Mes\sim\epsilon^{2/3}64\lambda^2\tau^{2a}\theta^{-\nu/2}\sqrt{2h}= 
\\
64\epsilon^{2/3} \lambda^2\sqrt{2f}\left(2a+1\right)^{-\frac{3a-b}{2(2a+1)}}\lambda^{-\frac{2b+3}{2(2a+1)}}
\times
\\
(\epsilon^{2/3}t)^{2a}\left(\frac{\lambda^2}{2a+1}(\epsilon^{2/3}t)^{2a+1}\right)^{-\frac{3a-b}{4a+2}}.
\end{eqnarray*}
After some calculations one obtains the formula for asymptotic behaviour of the measure, which depends on $t$ and $\varepsilon$:
\begin{equation}
Mes\sim 64\epsilon^{(2+a+b)/3} \sqrt{2\lambda f t^{a+b}}.
\label{AsymptoticsOfMesure}
\end{equation}

\section{Conclusions}

In this work were obtained a lot of new results. First of all we have found the conditions for chirp-rate and amplitude of driver for autoresonace. There are $a>0$ and $a-1<b<3a$.  This conditiona  are thresholds for the autoresonance.   Second result is the conditions for capture of the trajectory into autoresonance. There are additional condition for parameters of equation (\ref{generatedPrimryResonanceEq}): $a-b>0$ and $a+b\not=0$. Third result is a condition for the emission of  trajectory out of the resonance:  $a-b<0$ and $a+b\not=0$. Also one can see the measure of autoresonant asymptotics for nonlinear oscillator (\ref{AsymptoticsOfMesure}).

\end{document}